      [1999/12/01 v1.4c Il Nuovo Cimento]
\documentclass{cimento}

\usepackage{graphicx}  
\title{The optical spectra of the brightest mid-IR-selected galaxies}
\author{K. I.~Caputi\from{ins:x}\ETC,
        \atque
the zCOSMOS and SCOSMOS collaborations}
\instlist{\inst{ins:x} Institute of Astronomy, Swiss Federal Institute of Technology (ETH H\"onggerberg), 
Wolfgang-Pauli-Strasse 16, CH-8093, Z\"urich, Switzerland.
 }
\PACSes{\PACSit{98.62.Bj}
\PACSit{98.62.Bj}}
\begin{document}

\maketitle

\begin{abstract}
We present here some of the first results we have obtained on the study of the optical spectra of {\em Spitzer/MIPS} 24 $\rm \mu m$-selected galaxies  in the COSMOS field.  This is part of a series of studies we are conducting to analyse the optical spectral properties of mid-infrared (mid-IR) galaxies with different IR luminosities up to high redshifts. The results shown here correspond to the brightest $S_{24 \, \rm \mu m}>2 \, \rm mJy$ IR galaxy population at $z<1$.
\end{abstract}

\section{Introduction}

Since the discovery of an extragalactic component for the IR background a decade ago~\cite{ref:pug},
numerous studies have demonstrated the key importance of IR galaxies to reconstruct the history of
galaxy star formation and stellar mass assembly. Recent surveys conducted with the {\em Spitzer Space Telescope} have allowed 
for the resolution of more than 70\% of the mid and far-IR background light~\cite{ref:dole} and a quite detailed 
understanding of the evolution of IR galaxies with cosmic epoch, e.g.~\cite{ref:lefl,ref:cap06a,ref:cap06b,ref:cap07}.

The IR galaxy luminosity function has a strong positive luminosity evolution from redshifts $z=0$ to 1 (see~\cite{ref:lefl,ref:cap07}), 
implying that more luminous IR galaxies dominate the IR output with increasing redshift. Beyond $z\sim1$, basically all the IR light is produced by luminous and ultra-luminous IR galaxies (LIRGs and ULIRGs, respectively) with bolometric IR luminosities $L_{\rm bol.}> 10^{11} \, \rm L_\odot$. Active galactic nuclei (AGN) have some  significant  (but not yet well-determined) contribution to the IR background and govern the shape of the bright-end of the mid-IR LF at $z\sim2$~\cite{ref:cap07}.

The study of the optical spectra of IR galaxies can reveal several aspects of the physical conditions in which star formation and AGN activity take place. What is the evolutionary stage of the hosts of IR activity? What is the relationship between dust emission and the observed  reddening in optical bands? Does the IR phase accompany all the process of star formation or is it set at a particular time?
The large sample of zCOSMOS spectra~\cite{ref:lil07} being taken over the 2 deg$^2$  COSMOS field offers a perfect opportunity to address  these questions.

\section{Identifying mid-IR galaxies with zCOSMOS}

 The COSMOS survey~\cite{ref:sco07} is composed of multi-wavelength observations over 2 deg$^2$ of the sky at nearly equatorial latitudes. Among these observations, a large spectroscopic campaign is being carried out with the VIMOS spectrograph on the Very Large Telescope (VLT), namely the zCOSMOS survey~\cite{ref:lil07}.  The zCOSMOS survey consists of two parts: 1) zCOSMOS-bright, whose aim is to obtain 20,000 optical spectra for a magnitude-limited $I<22.5$ (AB mag) source sample and which is now half-way complete; 2) zCOSMOS-deep, which will yield 10,000 optical spectra in the central 1 deg$^2$ of the field, for galaxies colour-selected to be at high redshifts.
 
 The COSMOS field is also completely covered by deep {\em Spitzer} maps from 3.6 through 160 $\rm \mu m$~\cite{ref:san07}. These data have been obtained as part of two  {\em Spitzer}  Legacy programs in the observation cycles 2 and 3.

 We present some of the first results of a series of studies we are conducting to analyse the optical spectral properties of 24 $\rm \mu m$-selected galaxies in the COSMOS field. We particularly  focus here on the brightest IR galaxies ($S_{24 \, \rm \mu m}>2 \, \rm mJy$) found at redshifts $z<1$, whereas a complete study of the zCOSMOS-bright spectra of more than 600 $S_{24 \, \rm \mu m}>0.30 \, \rm mJy$ galaxies is presented elsewhere~\cite{ref:cap08}.

\section{24 $\rm \mu m$-selected galaxies in the COSMOS field: \\
the 2 mJy population}

 The completed first half of the zCOSMOS-bright program provides us with good-quality redshifts and spectra for 24 IR galaxies with $S_{24 \, \rm \mu m}>2 \, \rm mJy$ over 1.5 deg$^2$ of the COSMOS field. These galaxies are tracers of the most intense star-formation and/or  AGN activity and, within zCOSMOS-bright, 22 out of 24 are found at redshifts $z<1$. Among these 24 brightest IR galaxies with $I<22.5$ (AB mag), only 3 are identified with AGN from their optical spectra and/or X-ray emission.

\begin{figure}
\begin{center}
\includegraphics[width=12cm]{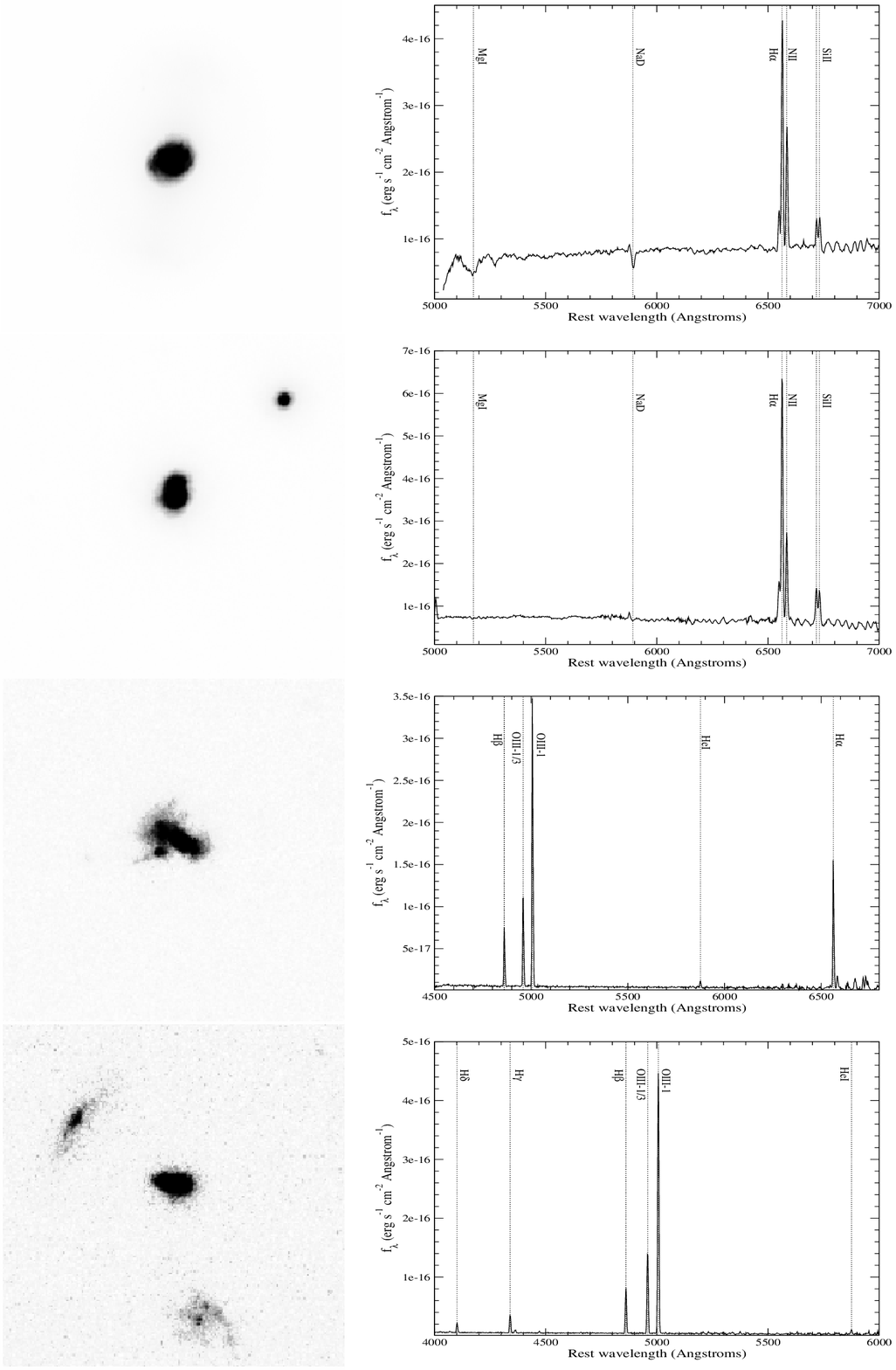}    
\caption{\label{fig1} {\em ACS} $I$-band postage stamps and zCOSMOS spectra for four galaxies with  $S_{24 \, \rm \mu m}>2 \, \rm mJy$ in the COSMOS field.  All these four galaxies are probably in the last stage of a merger.}
\end{center}
\end{figure}

 Figure \ref{fig1} shows {\em Advanced Camera for Surveys (ACS)} $I$-band postage stamps and zCOSMOS spectra for four of these IR sources with  $S_{24 \, \rm \mu m}>2 \, \rm mJy$.  The four  examples correspond to galaxies at redshifts $0<z<0.5$. In all these cases, the optical spectra display prominent emission lines, which are characteristic of actively star-forming galaxies. In fact, the intrinsic emission is even larger in all these galaxies, as all of them have at least mildly important degrees of dust obscuration, with $V$-band extinctions ranging   between $A_V<1$ and 2.5 (as obtained by equating the [IR+UV] and the optically-derived star formation rates).
 
 The {\em ACS} stamps of the four example galaxies reveal compact nuclear morphologies, where two different objects can be resolved in each case. These are probably galaxies in an advanced stage of a merger, as most of the brightest IR galaxies are thought to be.

\begin{figure}
\begin{center}
\includegraphics[width=8cm]{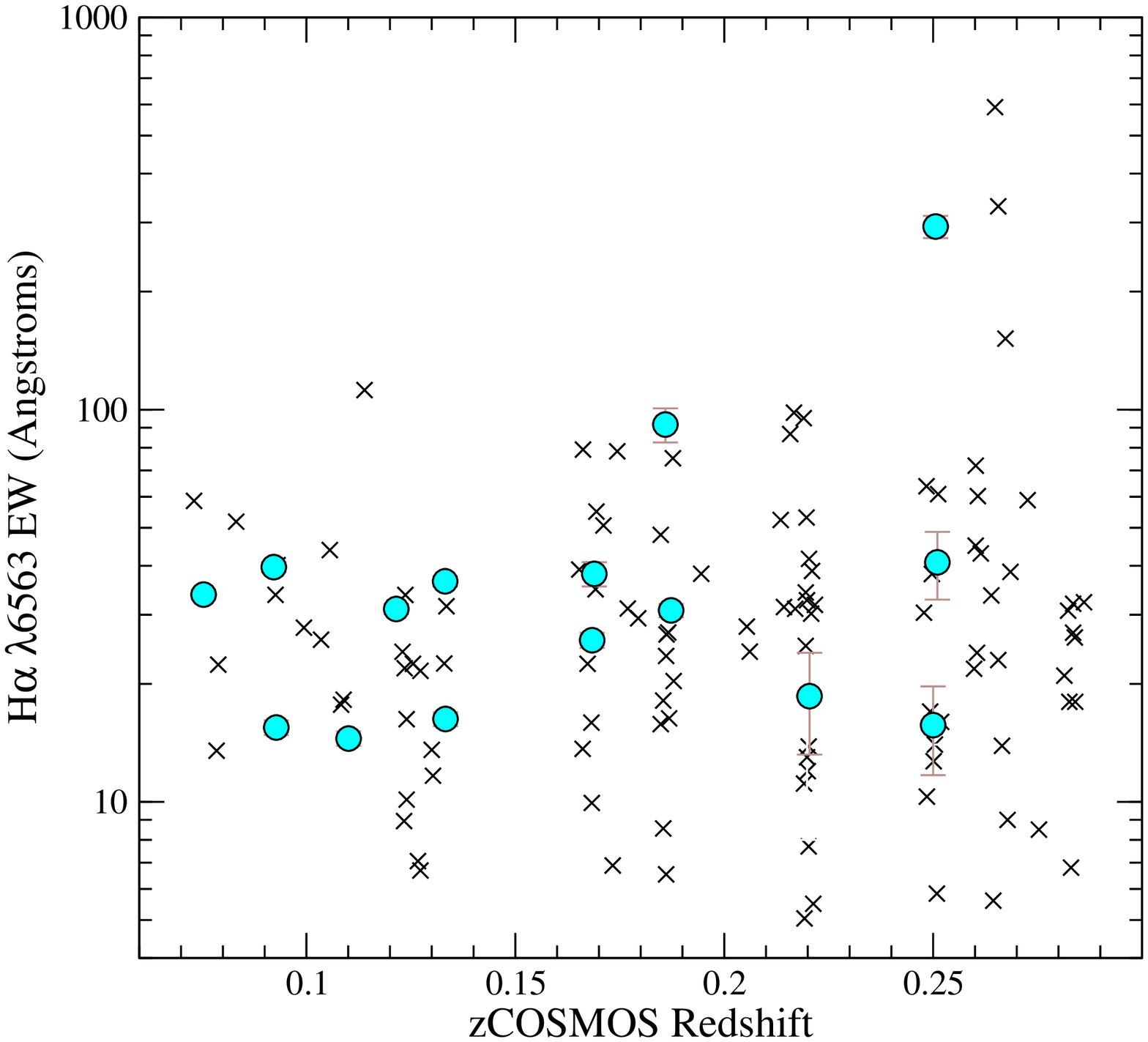}    
\caption{\label{fig2} The H$\alpha$ equivalent widths of the IR galaxies with $S_{24 \, \rm \mu m}>2 \, \rm  mJy$  as a function of redshift (filled circles), in comparison to  all the other galaxies with $S_{24 \, \rm \mu m}>0.30 \, \rm mJy$  (cross-like symbols).}
\end{center}
\end{figure}

\begin{figure}
\begin{center}
\includegraphics[width=8cm]{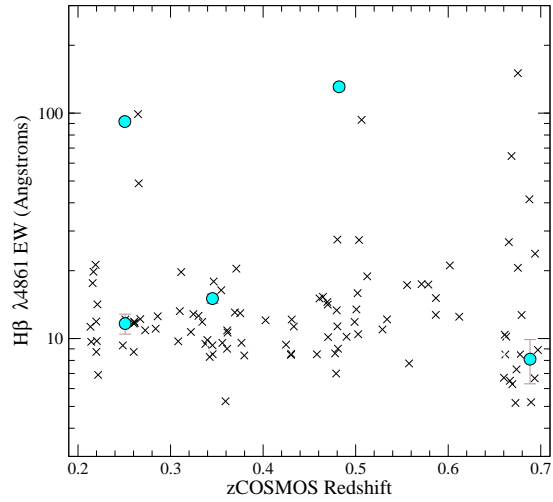}    
\caption{\label{fig3} The H$\beta$ equivalent widths of the IR galaxies with $S_{24 \, \rm \mu m}>2 \, \rm  mJy$ as a function of redshift, in comparison to  all the other galaxies with $S_{24 \, \rm \mu m}>0.30 \, \rm mJy$. Symbols are the same as in figure~\ref{fig2}.}
\end{center}
\end{figure}

 The equivalent widths of the emission lines provide information on the strength of the emission produced by the new generations of stars with respect to the underlying older stellar populations, which mostly contribute to the spectral continuum. Figures \ref{fig2} and \ref{fig3} show the H$\alpha$ and H$\beta$ equivalent widths, respectively, for the $S_{24 \, \rm \mu m}>2 \, \rm  mJy$ galaxies (filled circles), in comparison to the equivalent widths of all the other $S_{24 \, \rm \mu m}>0.30 \, \rm mJy$ sources (cross-like symbols). H$\alpha$ measurements have been performed for galaxies at $z<0.3$ and  H$\beta$ measurements correspond to galaxies at $0.2<z<0.7$.  The equivalent widths  are only shown  in the cases that they are larger than  5 $\rm \AA$.

Inspection of figures \ref{fig2} and \ref{fig3} shows that some of the brightest IR galaxies display very large H$\alpha$  or H$\beta$  equivalent widths. However, in many other cases, the equivalent widths are much smaller and indistinguishable of those of fainter IR galaxies. Thus, as other optical spectral characteristics~\cite{ref:cap08}, the Balmer line equivalent widths have little correlation with the IR photometric properties.

This simple comparison of equivalent widths illustrates the heterogeneity of the sources making the brightest IR-galaxy population at low redshifts.  The differences in equivalent widths are to some extent due to the different relative importance that the previously formed stars have in these galaxies.  Some of the brightest IR galaxies are relatively  young systems still in the process of chemical enrichment, while others are already-massive galaxies with strong stellar continua,  which might possibly be forming their last generations of stars~\cite{ref:cap08}.

\section{Future work}

With the availability of the deepest 24 $\rm \mu m$ images for the COSMOS field down to flux limits $S_{24 \, \rm \mu m} \sim 80 \, \rm  \mu Jy$, we will be able to explore the variations of optical spectral properties of mid-IR galaxies over a wide range  
of IR luminosities. On the other hand, by next year, zCOSMOS-deep will already  have produced one of the largest samples of optical spectra for galaxies at redshifts $z>1.5$. This will enable us to conduct an unprecedented study of the rest-frame UV spectral properties of IR  galaxies and to compare them to the spectra of other known high-$z$ galaxy populations, as e.g. Lyman-break galaxies.

\end{document}